\documentclass[namedreferences,hyperref,optionalrh]{spr-sola}
\usepackage{graphicx}        
\usepackage{color}           
\usepackage{amsmath}

\chardef\us=`\_

\begin{document}

\begin{frontmatter}
\title{The Solar eruptioN Integral Field Spectrograph}

\author[addressref={aff1,aff2},email={Vicki.Herde@LASP.Colorado.edu}]{\inits{V.L.}\fnm{Vicki~L.}~\snm{Herde}\orcid{0000-0001-9139-8939}}
\author[addressref={aff1,aff2},corref,email={Phil.Chamberlin@LASP.Colorado.edu}]{\inits{P.C.}\fnm{Phillip~C.}~\snm{Chamberlin}\orcid{0000-0003-4372-7405}}
\author[addressref={aff3},email={Donald.Schmit@Colorado.edu}]{\inits{D.}\fnm{Don}~\snm{Schmit}\orcid{0000-0002-9654-0815}}
\author[addressref={aff8},email={Adrian.Daw@NASA.gov}]{\inits{A.}\fnm{Adrian}~\snm{Daw}\orcid{0000-0002-9288-6210}}
\author[addressref={aff9},email={R.Milligan@qub.ac.uk}]{\inits{R.O.}\fnm{Ryan O.}~\snm{Milligan}\orcid{0000-0001-5031-1892}}
\author[addressref={aff4,aff10},email={Polito@lmsal.com}]{\inits{V.}\fnm{Vanessa}~\snm{Polito}\orcid{0000-0002-4980-7126}}
\author[addressref={aff4,aff5,aff6,aff7},email={Bose@lmsal.com}]{\inits{S.}\fnm{Souvik}~\snm{Bose}\orcid{0000-0002-2180-1013}}

\author[addressref={aff2},email={Spencer.Boyajian@lasp.colorado.edu}]{\inits{S.}\fnm{Spencer}~\snm{Boyajian}\orcid{0000-0002-4850-836X}}
\author[addressref={aff2},email={Paris.Buedel@lasp.colorado.edu}]{\inits{P.}\fnm{Paris}~\snm{Buedel}}
\author[addressref={aff2},email={William.Edgar@lasp.colorado.edu}]{\inits{W.}\fnm{Will}~\snm{Edgar}}
\author[addressref={aff1,aff2},email={Alex.Gebben@lasp.colorado.edu}]{\inits{A.}\fnm{Alex}~\snm{Gebben}\orcid{0009-0001-8920-1954}}
\author[addressref={},email={qian.gong-1@nasa.gov}]{\inits{Q.}\fnm{Qian}~\snm{Gong}\orcid{0000-0002-7187-1191}}
\author[addressref={aff2},email={Ross.Jacobsen@lasp.colorado.edu}]{\inits{R.}\fnm{Ross}~\snm{Jacobsen}}
\author[addressref={aff2},email={Nicholas.Nell@lasp.colorado.edu}]{\inits{N.}\fnm{Nicholas}~\snm{Nell}\orcid{0000-0001-7131-7978}}
\author[addressref={aff2},email={Bennet.Schwab@lasp.colorado.edu}]{\inits{B.}\fnm{Bennet}~\snm{Schwab}\orcid{0000-0002-1426-6913}}
\author[addressref={aff2},email={Alan.Sims@lasp.colorado.edu}]{\inits{A.}\fnm{Alan}~\snm{Sims}\orcid{0000-0002-4546-2394}}
\author[addressref={aff2},email={David.Summers@lasp.colorado.edu}]{\inits{D.}\fnm{David}~\snm{Summers}\orcid{0000-0003-0903-2212}}
\author[addressref={aff1,aff2},email={Zachary.Turner-1@colorado.edu}]{\inits{Z.}\fnm{Zachary}~\snm{Turner}}
\author[addressref={aff2},email={Trace.Valade@lasp.colorado.edu}]{\inits{T.}\fnm{Trace}~\snm{Valade}\orcid{0009-0009-5457-3991}}
\author[addressref={aff1,aff2},email={Joseph.Wallace@lasp.colorado.edu}]{\inits{J.}\fnm{Joseph}~\snm{Wallace}\orcid{0009-0008-2454-5212}}

\address[id=aff1]{University of Colorado Boulder, Boulder CO 80303}
\address[id=aff2]{Laboratory for Atmospheric and Space Physics, 3665 Discovery Dr, Boulder CO 80303}
\address[id=aff3]{Cooperative Institute for Research in Environmental Sciences,  216 UCB Boulder, CO 80309}
\address[id=aff4]{Lockheed Martin Solar \& Astrophysics Laboratory, Palo Alto, CA 94304, USA}
\address[id=aff5]{Bay Area Environmental Research Institute, NASA Research Park, Moffett Field, CA 94035, USA}
\address[id=aff6]{Institute of Theoretical Astrophysics, University of Oslo, PO Box 1029, Blindern 0315, Oslo, Norway}
\address[id=aff7]{Rosseland Center for Solar Physics, University of Oslo, PO Box 1029, Blindern 0315, Oslo, Norway}
\address[id=aff8]{Solar Physics Laboratory, NASA Goddard Spaceflight Center, 8800 Greenbelt Rd, Greenbelt, MD 20771}
\address[id=aff9]{Queen's University Belfast, University Rd, Belfast BT7 1NN, UK}
\address[id=aff10]{Department of Physics, Oregon State University, 301 Weniger Hall, Corvallis, OR 97331}

\runningauthor{Herde et al.}
\runningtitle{The Solar Eruption Integral Field Spectrograph}

\begin{abstract}
The Solar eruptioN Integral Field Spectrograph (SNIFS) is a solar-gazing spectrograph scheduled to fly in the summer of 2025 on a NASA sounding rocket. Its goal is to view the solar chromosphere and transition region at a high cadence (1~s) both spatially (0.5$''$) and spectrally (33~m\AA) viewing wavelengths around Lyman Alpha (1216\AA), Si~{\sc iii} (1206\AA) and O~{\sc v} (1218\AA) to observe spicules, nanoflares, and possibly a solar flare. This time cadence will provide yet-unobserved detail about fast-changing features of the Sun. The instrument is comprised of a Gregorian-style reflecting telescope combined with a spectrograph via a specialized mirrorlet array that focuses the light from each spatial location in the image so that it may be spectrally dispersed without overlap from neighboring locations. This paper discusses the driving science, detailed instrument and subsystem design, and pre-integration testing of the SNIFS instrument.
\end{abstract}
\end{frontmatter}

\section{Introduction and Background} 
     \label{S-Introduction} 

The lower solar atmosphere is temporally dynamic and spatially inhomogeneous, and it is becoming increasingly clear that this complex region must be better observed and studied if we are to fully understand how mass and energy are transported into the corona and beyond \citep{carlsson2019new}.

The Solar eruptioN Integral Field Spectrograph (SNIFS) is designed to break new ground by using a unique set of capabilities to probe the most vexingly complex region of the solar atmosphere: the chromosphere. One of the most important chromospheric lines, Hydrogen Lyman-alpha (Ly-$\alpha$; 1216\AA), is the brightest line in the solar ultraviolet (UV) spectrum. It is the dominant source of radiative energy loss from the lower transition region (TR) and forms over a wide range of temperatures and heights in the Sun \citep{fontenla1988lyman, fontenla2002energy}. 

Sounding rockets have a long tradition of observing Solar Ly-$\alpha$ since the beginning of the space age, due to strong atmospheric absorption preventing ground-based observations. \citealp{rense1953intensity} from the University of Colorado Boulder first observed the line using a grazing-incident spectrograph on an Aerobee rocket in 1952. This was followed by observations from \citealp{tousey1964sun} and many others taking advantage of the new capabilities sounding rockets provided. More recently, the Very High Angular Resolution Ultraviolet Telescope (VAULT 1.0, VAULT 2.0; \citealt{korendyke2000flight, vourlidas2016investigation}) rockets observed high-resolution Ly-$\alpha$ images while the Chromospheric Lyman-Alpha Spectro-Polarimeter instrument \citep[CLASP]{kano2012chromospheric}{} used Ly-$\alpha$ observations to explore the chromospheric magnetic field. 

Observations in Ly-$\alpha$ emission are ideally suited to fill in an observational gap in the crucial chromosphere/transition region (TR) interface since the spectral line is sensitive to temperatures ranging between 20~000--100~000K \citep{avrett2008models}. In addition, Ly-$\alpha$'s observations provide high signal-to-noise with little photospheric contamination due to the low UV continuum photospheric contamination at those wavelengths, which is critical for identifying spectral signatures of chromospheric events \citep{chintzoglou2018bridging}.

SNIFS will not only observe Ly-$\alpha$ but also Si~{\sc iii} (1206\AA, logT[K]=$4.65$) and O~{\sc v} (1218\AA, logT[K]=$5.35$), two TR lines that will allow us to explore how the chromosphere connects with the upper atmosphere. The SNIFS rocket mission has the primary objective to explore the energetics and dynamics of the chromosphere and transition region using a next-generation solar spectral imager, which is extensively described in this paper. SNIFS will be the first of its kind: a solar ultraviolet integral field spectrograph (IFS; \citealt{chamberlin2016integral}). 

This paper begins by discussing the scientific motivations behind the design of SNIFS and the current questions about solar energy transport that an instrument like SNIFS, combined with modeling, may be able to help answer. Section~\ref{S-ScienceSpicules} introduces spicules, which are SNIFS's primary observation target, and provides information on complimentary observing capabilities. Section~\ref{S-ScienceNanoflares} discusses nanoflares, with particular focus on modeling efforts which SNIFS will help constrain. Section~\ref{S-ScienceFlares} discusses current solar flare observation capabilities and and the lack of important high-cadence Ly-$\alpha$ observations, which SNIFS can provide. After this, we move on to the instrument design, starting by highlighting the new technologies which make SNIFS unique in Section~\ref{s-EnablingTechnologies}. Following this, we discuss the Optical (Section~\ref{S-OpticalDesign}), Mechanical (Section~\ref{S-MechanicalDesign}), Thermal (Section~\ref{S-ThermalCoolingSystem}), Electrical (Section \ref{S-electronics}), and Software (Section~\ref{S-SoftwareDataProcessing}) design. Finally Section~\ref{S-Testing} covers the Detector and Reflectance testing before wrapping up with a conclusion in Section~\ref{S-Conclusion}.


\subsection{Spicules}
    \label{S-ScienceSpicules}
 The chromosphere is a highly dynamic region, regularly exhibiting flows of the order of 50~km/s, despite being only a few thousand kilometers thick \citep{carlsson2019new}. It is dominated by a myriad of different features such as spicules, dynamic fibrils, and filaments, to name a few. Alfv\'enic and acoustic waves continuously traverse through it. In thermodynamic terms, it is highly heterogeneous: the mass density of the $\tau=1$ surface in chromospheric line cores can vary by several orders of magnitude in neighboring structures \citep{2007_rutten_chrom}. Moreover, the radiative transfer is dominated by non-local thermodynamic effects, which makes interpreting the observations challenging. The chromosphere plays a critical role in mass loading and heating the solar corona since all the nonthermal energy propagates through this region. Understanding the connection between the chromosphere and the corona is thus essential to model the coronal magnetic field accurately and to quantify the physical mechanism that heats the corona. One of the major scientific objectives of the SNIFS sounding rocket mission is to understand the impact of spicules in the upper chromosphere and the TR.
 
 Spicules are dynamic, thin, jet-like, ubiquitous features that permeate the solar chromosphere. Their importance in impacting the dynamics of the solar atmosphere has long been recognized \citep{1968SoPh....3..367B,beckers1972solar} ever since their discovery by Secchi in 1871 \citep{secchi1871atti}. Their presence in both active and quiet regions on the Sun makes their study exciting from the prospect of heating and mass loading of the solar corona \citep{1982_Athay_Holzer,depontieu2011origins}. Estimates suggest that there are more than a million spicules on the Sun at any given time \citep{beckers1972solar}. Over the past 20 years, a great deal of detail has emerged on their spectral and general characteristics thanks (primarily) to the development of high resolution ground-based and space-based telescopes that could achieve a spatial resolution better than 0.1$''$ \citep[e.g.][]{de2004solar,2007_bdp_spicules-typ1, VanDerVoort2009disk, sekse2012statistical,2013ApJ...767...17Y,  bose2019characterization, bose2021spicules, bate2022high,2023_bose_cbp}. 
 
 The interest in their study was massively revived after the advent of seeing-free, high-cadence observations were made possible by the Hinode spacecraft \citep{2007SoPh..243....3K}. Using off-limb observations in the Ca~{\sc ii}~H passband, \citealp{dePontieu2007tale} discovered a new class of spicules (type-II) which were far more dynamic, short-lived, and rapid compared to their classical (type-I) counterparts, and appeared to ``fade'' during their evolution. This rapid fading was perceived as a sign of heating and/or opacity changes during their evolution, which was resoundingly confirmed after the launch of the Interface Region Imaging Spectrograph \citep[IRIS;][]{DePontieu_2014} mission, where the fading Ca~{\sc ii}~H spicules subsequently appeared in the hotter Mg~{\sc ii} ultraviolet and (even) Si~{\sc iv} passbands \citep{2014_Tiago_spicules}. IRIS has since provided high-resolution spectroscopic and imaging observations in the chromospheric (Mg~{\sc ii} ultraviolet doublet) and TR (Si~{\sc iv}) channels, which provides rich thermal and velocity diagnostics \citep{skogsrud_15, bose2019characterization, 2020_don_intensity_fluc, 2021A&A...654A..51B, herde2023spicules}. 
 
 
The formation temperature range of the Ly-$\alpha$ spectral line \citep[20~000--100~000K,][]{avrett2008models} makes it an ideal candidate to study the impact of spicules in the upper chromosphere and the TR with unprecedented detail. This serves as an important link between partially and fully ionized plasma in the chromosphere and the corona. However, there are only a handful of investigations associated with spicules in the Ly-$\alpha$ channel. Previous studies have focused on spicules and/or their associated intensity disturbances from a spectroscopic \citep[using CLASP,][]{2019_clasp_yoshida_wave} and imaging \citep[using CLASP and VAULT2.0,][]{2016_CLASP_Kubo,2018_georgios_vault2,2020_don_intensity_fluc} perspective. However, the relatively lower spatial ($\sim$3$''$) and spectral (100m\AA) resolution of CLASP instrument \citep{2016SPIE.9905E..3DG} allowed a spectroscopic investigation of only 1 spicule over its entire duration of observation \citep{2019_clasp_yoshida_wave}. Its high cadence (0.3s) imaging capability did result in the discovery of upper chromospheric/lower TR rapid intensity fluctuations \citep{2016_CLASP_Kubo,2017_CLASP_Ishikawa}, but there was no spectral information possible. VAULT2.0 observations, on the other hand, did have much higher spatial resolution \citep[$\sim$0.5$''$,][]{2016_VAULT2_vourlidas} but owing the relatively lower cadence ($\sim$8s) it was only possible to study one type-II spicule in detail \citep{2018_georgios_vault2}. SNIFS has the best of both worlds. Its high cadence and spectrally resolved measurements of the brightest spectral line in the UV will provide critical insight into the origins, dynamics, and propagation of fast-changing fine-scale structures, like spicules, in chromospheric and transition region spectra over a two-dimensional field of view simultaneously. Its Si~{\sc iii} and O~{\sc v} observations will also allow these events to be traced lower into the chromosphere and higher into the transition region. This is possible due to new advances in EUV integral field spectroscopy (described in Section~\ref{s-EnablingTechnologies}), which will provide resolved multi-line spectral measurements and 2D spatial imaging simultaneously. SNIFS will allow us to analyze many spicules in a single flight dataset, from formation to disappearance.

\subsection{Nanoflares}
    \label{S-ScienceNanoflares}
   
Flares are dramatic explosive events in the solar atmosphere, associated with the release of energy in various forms, including emission over a broad range of wavelengths, particle acceleration, and magnetohydrodynamic (MHD) waves \citep{Fletcher2011}. 
A small flare in a typical active region will generate $\approx$10 times the amount of soft X-ray flux in a 100 square Mm patch than the rest of the Sun combined. It has been suggested that the relationship between flare power and event frequency follows a power-law distribution \citep[e.g.][]{Hudson1991,hannah2008rhessi}. Small but ubiquitous energy releases, of the order of $\approx$10$^{24}$ ergs (faint, as-yet unresolvable energetic events predicted by \citealt{1988ApJ...330..474P}) may thus be responsible for heating the corona to million degrees Kelvin. 
Several recent studies suggest that smaller microflare or nanoflare-type size events found in active regions may, in fact, behave as scaled-down versions of larger flares, where the plasma may reach temperatures up to $\sim10$~MK \citep{reale2019large, glesenerAcceleratedElectronsObserved2020a, testaCoronalTemperatureSolar2020a, testaIRISObservationsShortterm2020a, cooperNuSTARObservationMinuscule2020a}, and particle acceleration may occur \citep{hannah2008rhessi, testaEvidenceNonthermalParticles2014a, wrightMicroflareHeatingSolar2017, testaIRISObservationsShortterm2020a, glesenerAcceleratedElectronsObserved2020a, cooperNuSTARObservationsRepeatedly2021}. 
The observation of widespread, faint emission in high-temperature lines Ca~{\sc xvii} and Fe~{\sc xix} (typically formed $\sim$6~MK) from quiescent active regions has been observed by \citealp{testa2012hinode} and \citealp{brosius2014pervasive}, respectively, and interpreted as due to unresolved nanoflares, but evidence of individual events is lacking.

In large flares, signatures of electron beam acceleration are typically observed by hard X-ray instruments \citep[e.g.][]{Krucker2008}. 
As predicted by the solar flare models, such electrons are accelerated from the reconnection region in the corona towards the lower atmosphere along the reconnected magnetic field lines. The accelerated particles are then stopped by the dense plasma in the chromosphere, producing unique and telltale signatures in chromospheric and TR line profiles. \citealp{testaEvidenceNonthermalParticles2014a}, \citealp{polito2018investigating}, and \citealp{testaIRISObservationsShortterm2020a} recently studied the chromospheric response to weak electron beams for smaller micro- or nanoflare-type events. 
These studies have shown that we can provide new indirect diagnostics of the presence of difficult-to-measure nonthermal particles in small heating events by comparing high-resolution observations of footpoint brightenings in TR and chromospheric lines from IRIS to state-of-the-art flare models.
In particular, the spectral characteristics of the emission lines are very sensitive to the mechanisms of energy transport (e.g., thermal conduction vs. nonthermal particles), as well as the details of the electron distribution (e.g., low energy cut-off, energy flux), providing tight constraints on the models. 

SNIFS will provide unprecedented new observations of small energetic heating events such as nanoflares and microflares in the largely unexplored Ly-$\alpha$ emission, as well as Si~{\sc iii} and O~{\sc v}. SNIFS observations, combined with modeling, have thus the potential to provide crucial diagnostics of heating mechanisms in nanoflare-type events. 

\begin{figure}[!htb]
    \centering
    \includegraphics[width=\textwidth]{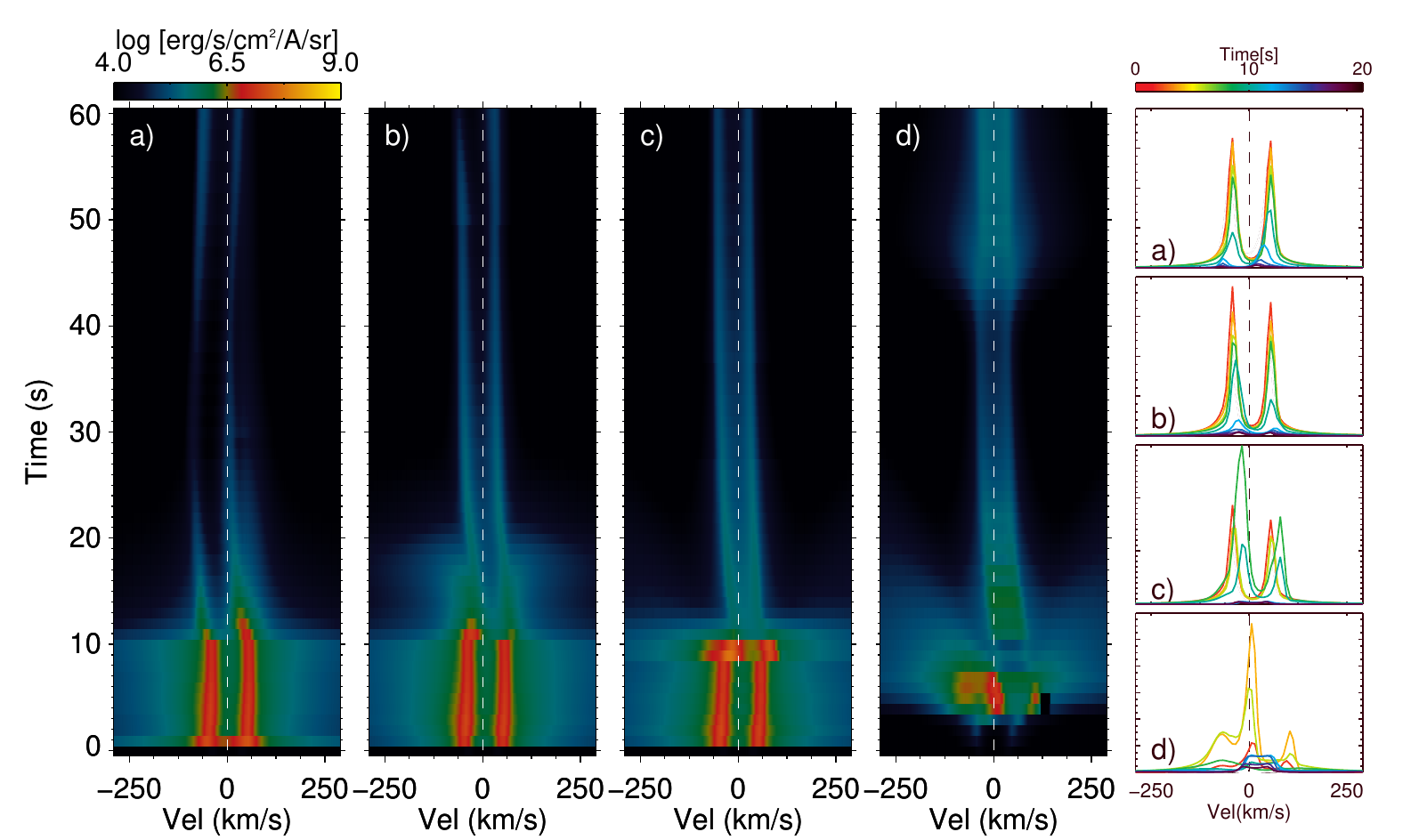}
    \caption{Synthetic Ly-$\alpha$ spectra as a function of time (a-d) as forward modeled using the RADYN and RH1.5D codes for nanoflare simulations with 10s heating by electron beams with energy cut-off E$_C$ of 8~keV and energy flux of 5$\times$10$^{8}$ (a) 8$\times$10$^{8}$ (b) and 1.2$\times$10$^{9}$ (c) ergs cm$^{-2}$ s$^{-1}$; as well as thermal conduction only (d) for a coronal loop with half-length L/2=15~Mm and initial loop top coronal temperature of 1~MK. Negative (positive) velocities indicate blue (red)-shifts. The panels on the right side show Ly-$\alpha$ spectra every 1s during the first 20s of each simulations. }
      \label{Fig:sims}
\end{figure}

An example is provided in Figure~\ref{Fig:sims}, showing synthetic Ly-$\alpha$ emission, as will be observed by SNIFS, for different nanoflare models and as a function of time and velocity.
The nanoflare simulations were performed using the RADYN code \citep{carlssonNonLTERadiatingAcoustic1992, carlssonDoesNonMagneticSolar1995, carlsson1997formation, allredRadiativeHydrodynamicModels2005, allredUNIFIEDCOMPUTATIONALMODEL2015}, which solves the coupled equations of hydrodynamics, radiative transfer and charge and level population conservation on a 1D flux tube rooted in the photosphere and stretching out to include the chromosphere, TR and corona. RADYN utilizes multiple heating mechanisms including: beams of accelerated particles traveling from the corona toward the chromosphere characterized by a low energy cut-off, E$_C$, and power law index, $\delta$; local heating of the corona, with subsequent transport of the energy by a thermal conduction front; and dissipation of Alfv\'{e}n waves \citep{kerrSIMULATIONSMGIi2016}.

Panels a) to c) show electron-beam heating models with a low energy cut-off (E$_C$) of 8~keV, spectral index equal to 10, and energy flux of 5$\times$10$^{8}$, 8$\times$10$^{8}$ and 1.2$\times$10$^{9}$~ergs~cm$^{-2}$~s$^{-1}$, respectively, while panel d) shows results for a thermal conduction model with equivalent energy flux of 1.2$\times$10$^{9}$~ergs~cm$^{-2}$~s$^{-1}$. The heating duration is in all cases 10s. The flare loop has a half-length of 15~Mm and an initial apex temperature of 1~MK. Negative (positive) velocities indicate blue (red) shifts of the line. The electron beam models are constrained based on a recent study combining IRIS observations with HXR observations from NuSTAR \citep{politoMultiwavelengthObservationsModeling2023}. Finally, the multiplot panels on the right side show Ly-$\alpha$ individual spectra every 1s during the first 20s of the simulations. 

Figure~\ref{Fig:sims} demonstrates that the spectral characteristics (Doppler shift at the line core, broadening, peak asymmetries, line profiles, etc.) of the Ly-$\alpha$ line are sensitive to the details of the heating mechanism. In real observations, we expect the heating to be highly dynamic and vary both in space and time. SNIFS will provide high-cadence observations with adequate spatial coverage that are crucial for distinguishing between possible competing nanoflare models.

A key strength of RADYN is the ability to model the important elements for the chromospheric energy balance (i.e., H, He, Ca) in non-LTE radiative transfer. Other chromospheric species are included as background continuum opacity sources in LTE. 
RADYN models emission lines and continua assuming complete redistribution (CRD). However, it is well known that the accurate model of some chromospheric lines (such as Ly-$\alpha$) requires including the effects of partial redistribution (PRD; \citealt{leenaartsFORMATIONIRISDIAGNOSTICS2013, leenaartsFORMATIONIRISDIAGNOSTICS2013a}). The synthetic profiles in Figure~\ref{Fig:sims} were obtained by feeding snapshots of the atmospheric temperature, density, and velocity stratification from the RADYN simulations into the time-independent radiative transfer RH 1.5D code \citep{uitenbroekMultilevelRadiativeTransfer2001, pereiraRH5DMassively2015} which allows us to take into account the effects of PRD. 

In preparation for the SNIFS launch, we will perform a larger parameter space study assuming different nanoflare heating models. By comparing in detail the observed and synthetic SNIFS Ly-$\alpha$ line profiles, also combined with the other TR lines (e.g., SI~{\sc iii}, O~{\sc v}), physical constraints on the nanoflare models will be derived.

\subsection{Solar Flares}
    \label{S-ScienceFlares}

The study of solar flares in Ly-$\alpha$ emission has been overlooked in recent years as attention has focused on EUV emission from the corona during these energetic events. Many of the recent Ly-$\alpha$ instruments have not had the cadence, sensitivity, or duty cycle to capture the fleeting impulsive phase in this fundamental line. Of the studies that do exist in the literature, the vast majority are based on spatially and spectrally integrated (photometric/irradiance) observations (e.g. SORCE/SOLSTICE \citep{rottman1993solar, mcclintock2005solar, snow2005solar}, SDO/EVE \citep{woods2012extreme}, GOES/EUVS \citep{eparvier2009extreme}, PROBA2/LYRA \citep{hochedez2006lyra, dominique2013lyra}, and MAVEN/EUVM \citep{jakosky2015mars, eparvier2009extreme}). 

Ly-$\alpha$ is known to radiate a significant fraction of the energy deposited in the chromosphere during solar flares \citep{milligan2014} and a study of almost 500 M- and X-class flares \citep{milligan2020lyman} observed by GOES-15/EUVS-E showed that Ly-$\alpha$ can radiated up to 100 times more energy than the more commonly observed X-rays. \citealp{greatorex2023lyman} also showed that flares of comparable magnitudes, occurring at similar locations on the solar disk (to negate any center-to-limb variation), can have strikingly different Ly-$\alpha$ responses. It is also currently unclear how much of a contribution coronal Ly-$\alpha$ emission makes to photometric flare observations, and whether such emission would be optically thick or thin. Studies by \citealp{milligan2021lyman} and \citealp{wauters2022} have suggested that this contribution may be significant even for modest events.

The only existing high-resolution Ly-$\alpha$ flare spectra were taken with the Naval Research Laboratory (NRL) spectrograph on the Apollo Telescope mount onboard Skylab in the 1970s \citep{canfield1980observed} although its data were not flux calibrated as standard\footnote{The SUMER instrument on SOHO was not permitted to be pointed at flaring regions due to the sensitivity of its optics}. The Ly-$\alpha$ line profile can be a useful diagnostic of plasma conditions under different heating conditions, particularly when compared with radiative hydrodynamic models \citep{kennedy2015radiative,brown2018modeling,brown2019observations}. As with nanoflares and spicules, SNIFS's wavelength range will allow the flaring solar atmosphere to be sampled over a range of heights, although for stronger heating scenarios, shedding light on the energy transport processes at work. Its imaging capability will also allow chromospheric and coronal contributions to be disentangled. 
Similar to the nanoflare investigation, the RADYN models described in Section~\ref{S-ScienceNanoflares} will also be used to help interpret SNIFS flare observations. 

Observing the impulsive phase of a solar flare using rocket-based instruments has historically been challenging, but SNIFS is part of the Solar Flare Sounding Rocket Campaign which has been working to develop flare forecasting capabilities for triggered observations. The Focusing Optics X-ray Solar Imager (FOXSI-4; \citealt{buitrago2021foxsi, glesener2022high}) and the High-Resolution Coronal Imager (Hi-C Flare; \citealt{kobayashi2014high}) rockets successfully launched to view a M~1.6 class solar flare on 17 April 2024. The flare alert software used for this campaign (Peterson et al., in preparation) will also be used by SNIFS to assist with launch decision-making. 


\section{Instrument Design} 
    \label{L-InstrumentDesign}
    
The SNIFS instrument is comprised of two sections: a fore section which houses most of the telescope elements, and an aft section which houses the spectroscope elements. A third section on the rocket houses electronics. The telescope portion is a standard Gregorian design with off-axis parabola primary, secondary, and tertiary mirrors, folding the light beam to a focus behind the primary mirror. At this focus, the instrument transitions to the IFS portion of the instrument. At the focal plane are placed two mirrorlet arrays, each of which is a planar array of microscopic concave mirrors etched into a single piece of glass. These mirrorlet arrays act similarly to a lenslet array such as the ones employed by the Spectroscopic Areal Unit for Research on Optical Nebulae (SAURON) instrument on the William Herschel Telescope \citep{miller2000sauron} and the Spectro-Polarimetric High-contrast Exoplanets Research (SPHERE) IFS on the Very Large Telescope (VLT; \citealt{claudi2010sphere}). The mirrorlets divide the large incoming beam of light into smaller beams of light. The two arrays each pick out a different field of view (FOV) in order to view both a flare footpoint and an area with solar spicules. The spectrograph elements include a focusing mirror, a grating, and a detector for each field of view.

\begin{figure}
\centerline{\hspace*{0.015\textwidth}
         \includegraphics[width=1\textwidth,clip=]{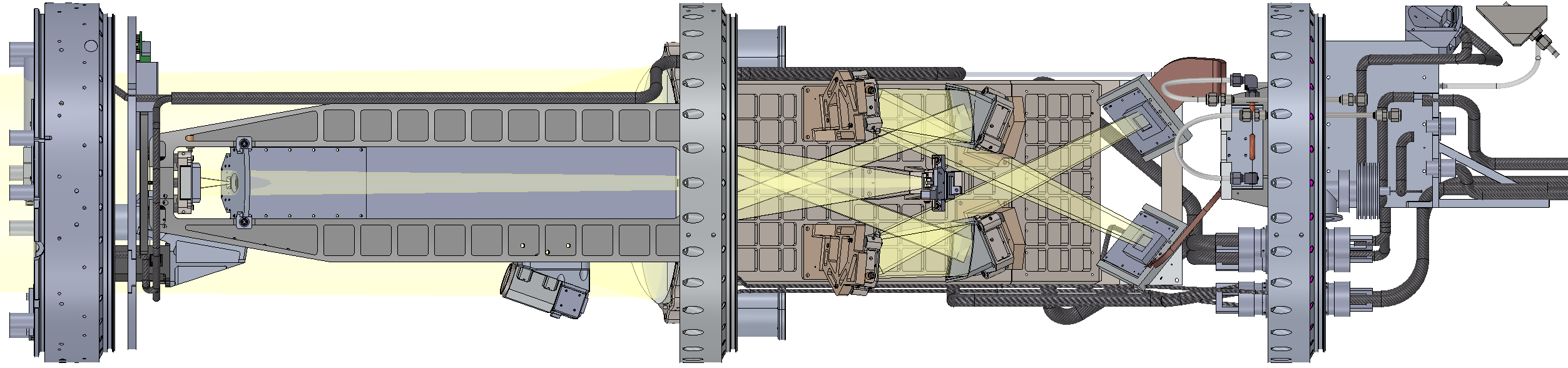}
        }
\vspace{-0.05\textwidth}   
\centerline{\Large \bf     
\hspace{0.90 \textwidth}  \color{black}{(a)}
   \hfill}
\centerline{\hspace*{0.015\textwidth}
         \includegraphics[width=1\textwidth,clip=]{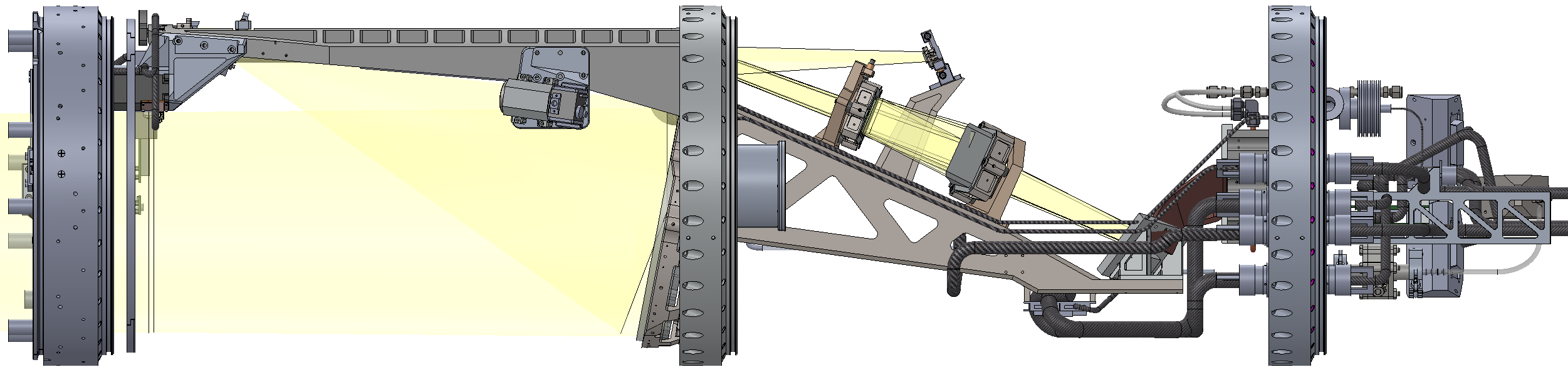}
        }
\vspace{-0.05\textwidth}   
\centerline{\Large \bf     
\hspace{0.9 \textwidth} \color{black}{(b)}
   \hfill}              
\caption{The SNIFS Instrument layout. (a) A top view of the payload. (b) A side view of the payload. The left two sections contain the telescope and spectrograph elements, while the right section contains telemetry and housekeeping elements. Light enters through a shutter door on the left of the payload. The primary bulkhead is located center left of the payload with optical benches mounted to it. The telemetry bulkhead is located center right on the payload carrying the large heatsink, electrical, and vacuum equipment.}
\label{F-Rocket_Layout}
\end{figure}

The SNIFS instrument is designed to fit within a 17.26-inch diameter sounding rocket payload skin. All optical elements either mount directly to the primary bulkhead or to the optical benches mounted on the primary bulkhead. The front two sections allow air to pass between them, but as a unit are hermetically sealed and can be drawn down to vacuum during testing. Behind these two sections is a telemetry section where most of the electrical boards and harnessing are stored.

\subsection{Enabling Technologies}
\label{s-EnablingTechnologies}
    \subsubsection{The Mirrorlet}

Table \ref{IFS_Technology} shows that IFS technology will be massively beneficial to solar physics by obtaining information in four dimensions simultaneously: two spatial dimensions, one spectral dimension (which, with the assistance of modelling, can provide both altitude and Doppler velocity), and high-cadence temporal dimension.

\begin{figure}[!tbp]
      \centering
      \begin{minipage}[b]{0.45\textwidth}
        \includegraphics[width=\textwidth]{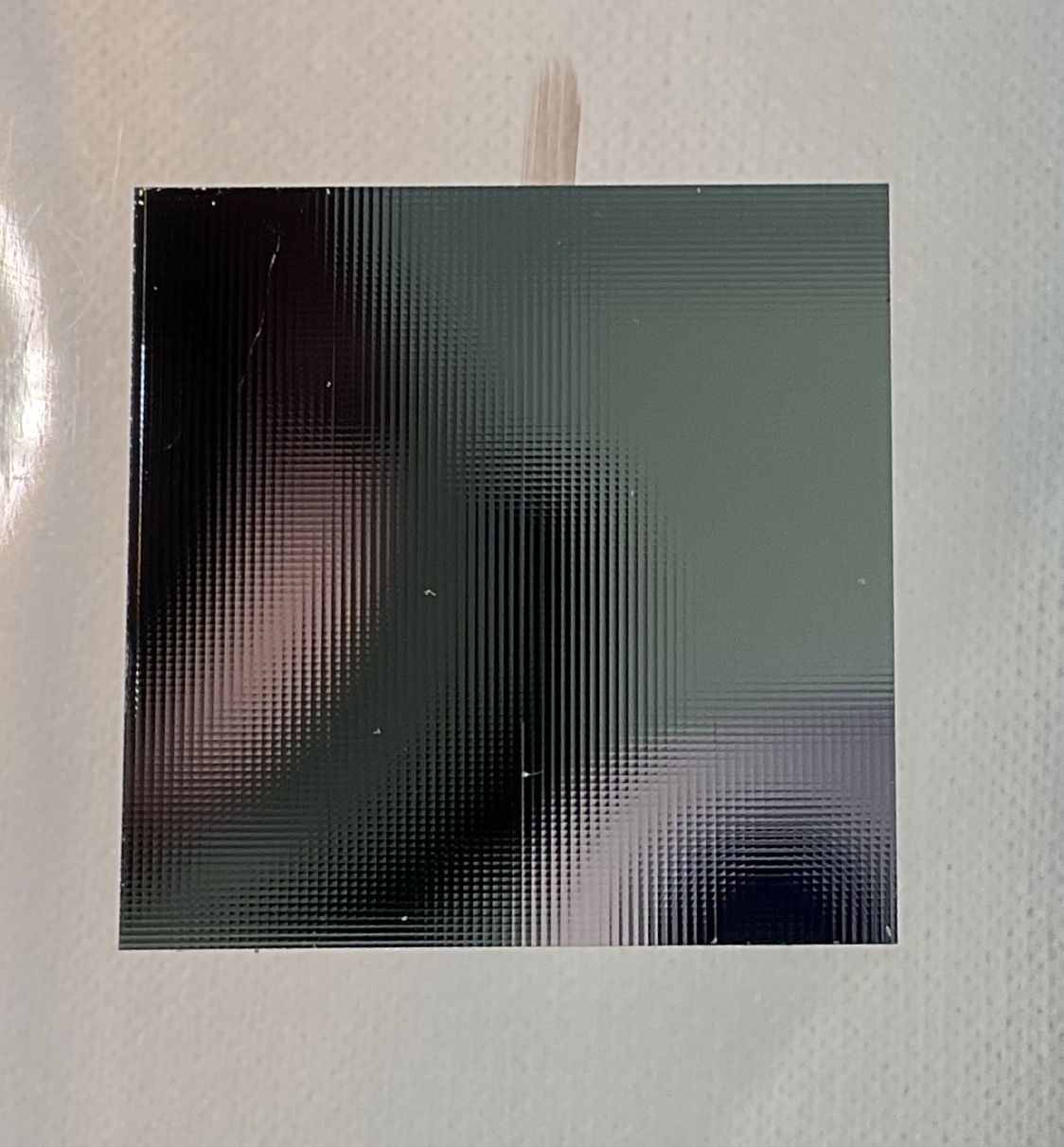}
        \put(0,0.1){(a)}
      \end{minipage}
      \hfill
      \begin{minipage}[b]{0.45\textwidth}
        \includegraphics[width=\textwidth]{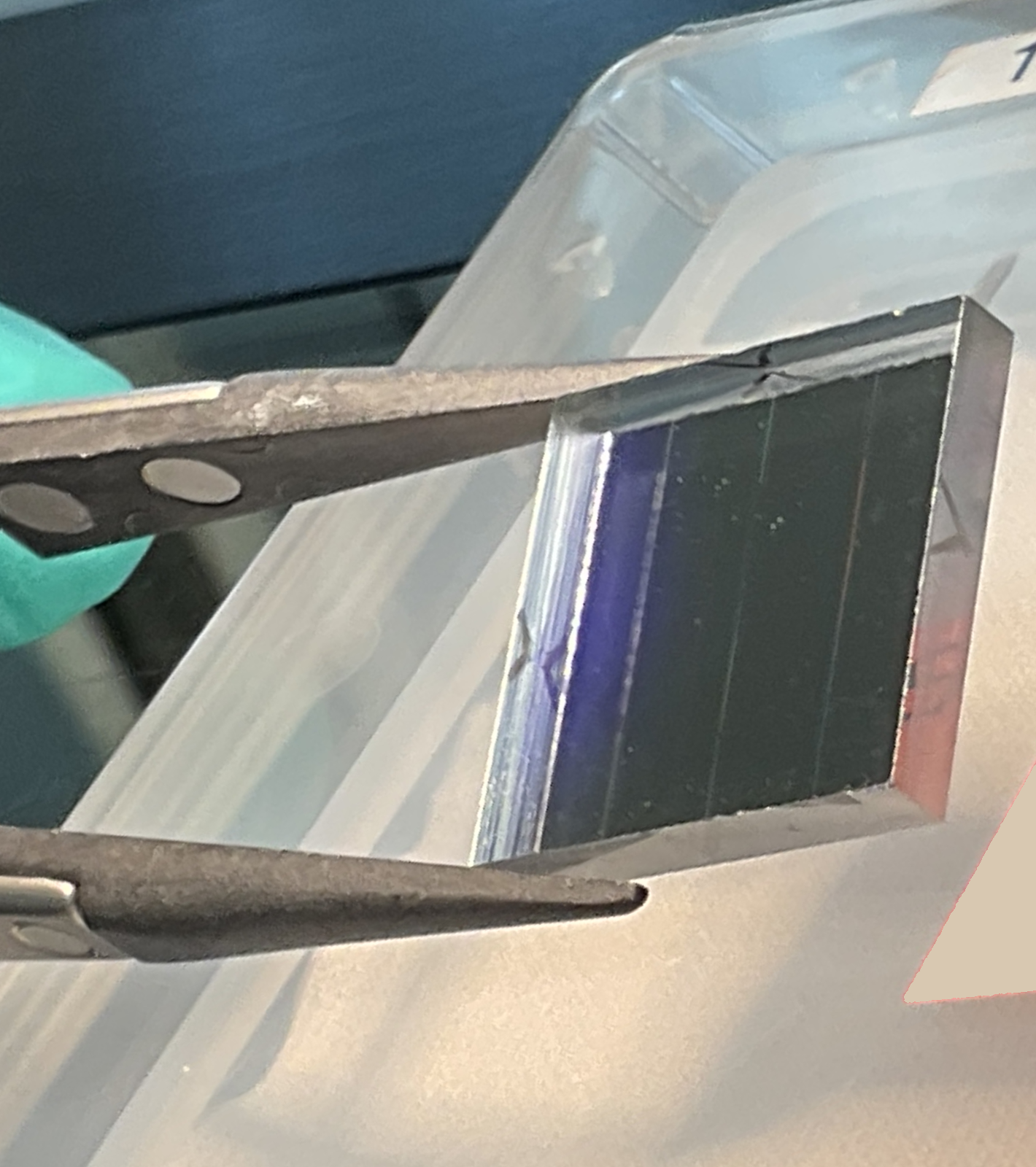}
        \put(0.25,0.1){(b)}
      \end{minipage}
      \caption{Photos of one of the mirrorlet arrays. (a) shows a view of the array's face. The individual mirrorlets can be seen as a reflective grid. (b) shows the back of the mirrorlet array, with tweezers for scale.}
      \label{F-mirrorlet_photo}
    \end{figure}

The primary technology which will allow SNIFS to take spectral images of the Sun at a 1s time cadence is a mirrorlet array developed by Dr. Phil Chamberlin and Dr. Qian Gong at GSFC \citealp{chamberlin2016integral}. This is a 80$\times$100 element 2D array of reflecting and focusing mirrorlets coated with Aluminum and MgF$_2$ (seen in Figure \ref{F-mirrorlet_photo} and manufactured by Jenoptik) to allow IFS capabilities to move down to shorter wavelengths. The mirrorlet array is placed at the imaging plane of the telescope, similar to the location of a slit in a traditional imaging slit spectrometer design. After the mirrorlet in the optical path, a focusing element and grating complete the spectrograph to produce a high-resolution spectrum for each spatial element defined by the mirrorlet elements.

    \begin{table}[!htb]
        \begin{center}
    	\includegraphics[width=1\textwidth]{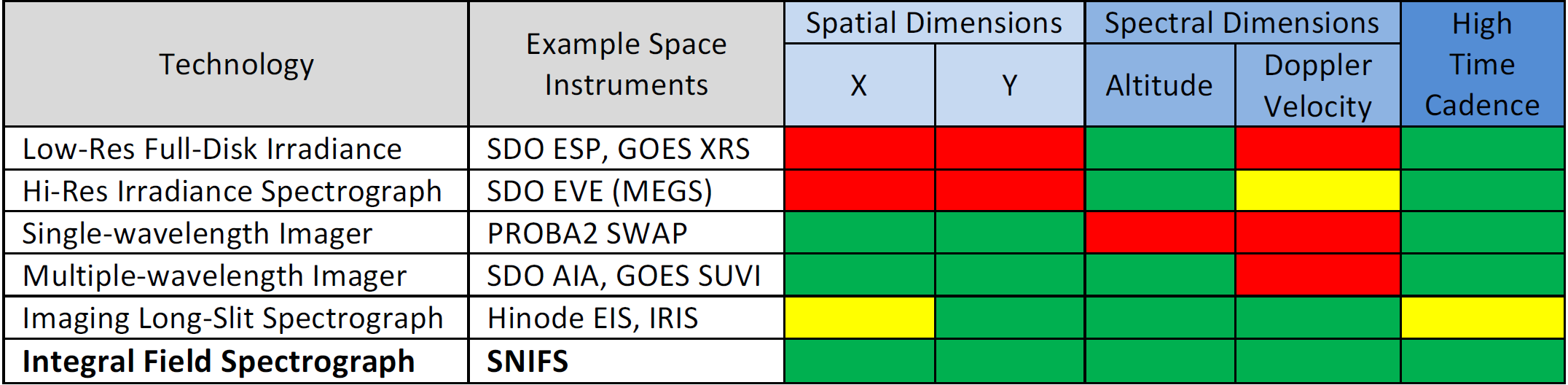}
        \end{center}
        \caption{Comparison of Technology for Spectral Imaging}
        {\raggedright Green is excellent capability, Yellow is limited or compromised capability, and Red is no capability. \par}
    \label{IFS_Technology}
    \end{table}
    
      \begin{figure}[!htb]
        \caption{A simplified version of how a mirrorlet array works, showing how an image at a focal plan is converted into spectra and back into a datacube. Adapted from \citealp{westmoquette2009integral}.}
        \begin{center}
        \begin{picture}(2,9)
            \put(-0.36\textwidth,0){(a)}
            \put(-0.1\textwidth,0){(b)}
            \put(0.1\textwidth,0){(c)}
            \put(0.35\textwidth,0){(d)}
          \end{picture}
    	\includegraphics[width=1\textwidth]{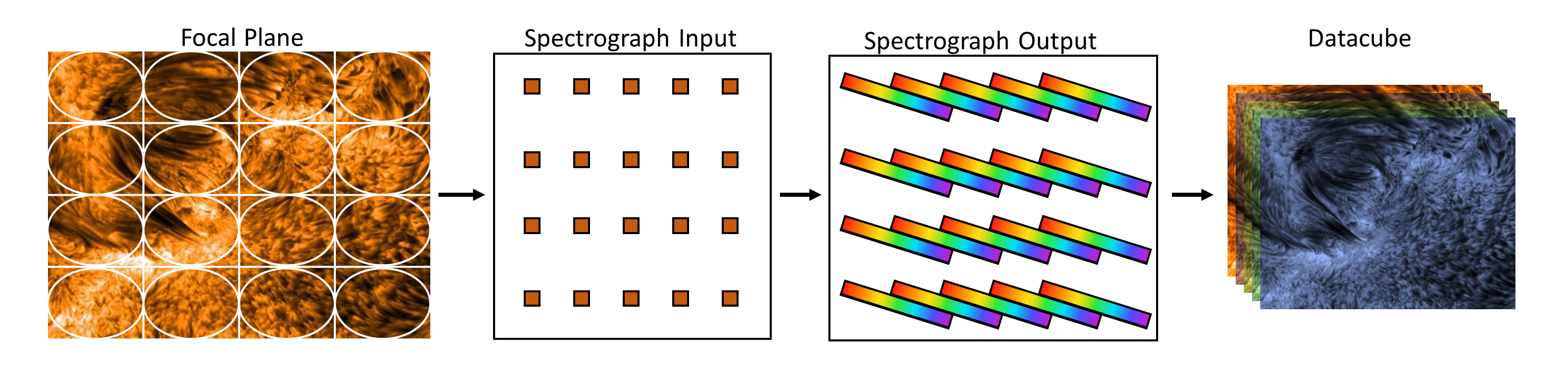}
        \end{center}
    \label{F-mirrorlet}
    \end{figure}
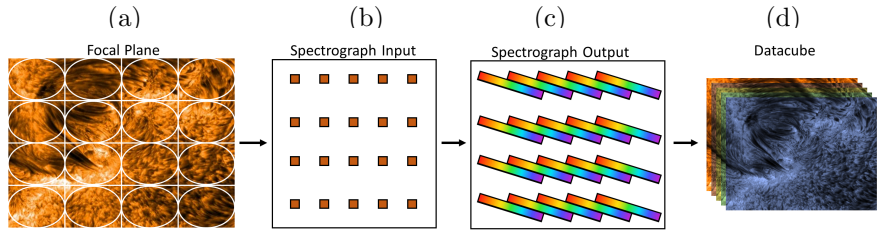

    Figure~\ref{F-mirrorlet} (adapted from \citealp{westmoquette2009integral}) shows a simplified cartoon of the spectrograph. Before reaching the mirrorlet array, a telescope focuses an image. A mirrorlet array made of a single planar piece of glass with hundreds of small concave mirrorlets etched into its surface is placed at the focal plane of the telescope (\ref{F-mirrorlet}a). The array of mirrorlets divides the image into an array of light beams, with each beam corresponding to one spatial pixel of the final image. The mirrorlets do not direct the rays to different locations as an image slicer would, but instead keep the rays in the same position relative to each other (\ref{F-mirrorlet}b). The array of light beams is then reflected off a holographic grating (focusing element not shown), turning the array of light beams into an array of spectra (\ref{F-mirrorlet}c). The grating and thus the spectra are rotated slightly to reduce spectral overlap. The spectra are then recorded by a detector at the focal plane of the system (\ref{F-mirrorlet}d). Because the array of spectra are kept in the same position relative to each other, the top left spectrum corresponds to the top left spatial pixel of the image, the bottom right spectrum to the bottom right pixel of the image, and so on. In the SNIFS payload, the spectra's lengths are constrained by a bandpass filter earlier in the telescope. This prevents excessive spectral overlap of neighboring pixels.

    \subsubsection{The Detector}
    The other new major enabling technology in the payload is the CSIE4K CMOS detector currently in development at the Laboratory for Atmospheric and Space Physics (LASP). This is a back-illuminated CMOS sensor, the Teledyne E2v CIS120, with low readout noise and dark current at 0~C. The image sensor and read electronics are capable of capturing and compressing up to 10 images per second while only consuming around 4W of power. Large CMOS detectors can have readout times orders of magnitude better than a similarly sized CCD, which means the SNIFS imaging cadence is dictated by science requirements, not detector readout time. Additional details about the detector design can be found in Section \ref{S-DetectorDesign}, while testing details are given in Section \ref{S-DetectorQEtesting}.

\subsection{Optical Design}
    \label{S-OpticalDesign}

\begin{table}
\centering
\begin{tabular}{ll}     
 \hline                   

Telescope EFL at mirrorlet array &	82.5 m \\
f/\# at mirrorlet array& 	273 \\
f/\# at CCD&	 40 \\
FOV	(x2)  & 0.5$\times$0.5 arc minute \\
Wavelengths &	120.5–-122.0 nm \\
Spatial resolution*&	0.6$''\times0.5''$/mirrorlet pix \\
Spectral resolution*& 33~m\AA \\
Grating period &	 2900 l/mm (2nd order)\\
Mirrorlet array & 80$\times$100 with 250$\times$200 $\mu$~m pixels\\
CMOS Imaging Sensor & 2048$\times$2048 with 10 $\mu$~m pixels \\

  \hline
\end{tabular}
\caption{SNIFS optical design parameters. }
{\raggedright * Resolutions listed are are the design ideal. Measured resolutions will be provided in in a later paper. \par}
\label{CutoffTable}
\end{table}

The optical design includes two main parts: the telescope assembly and the IFS assembly.

The telescope was designed to have an 82.5~m focal length and a large f/\# of 273 to meet the spatial magnification requirements for the mission.  A three-mirror telescope structure was used in the design: the first two mirrors create a Gregorian type of telescope, while the third mirror extends the focal length to 82.5~m to meet the spatial resolution on each mirrorlet pixel of 0.5$''$/mirrorlet pixel while also folding the design to fit more compactly within the rocket. A Gregorian design was chosen so that a baffle could be inserted at the focus of the primary mirror before the secondary mirror for the purpose of light and heat rejection. Concentrating sunlight from an 11$''$ diameter beam to a $<$1$''$ diameter beam is likely to heat up and damage reflecting mirror elements. Refer to Section~\ref{S-Baffles} for steps taken to mitigate this.


\begin{figure}
\centerline{\hspace*{0.015\textwidth}
         \includegraphics[width=1\textwidth,clip=]{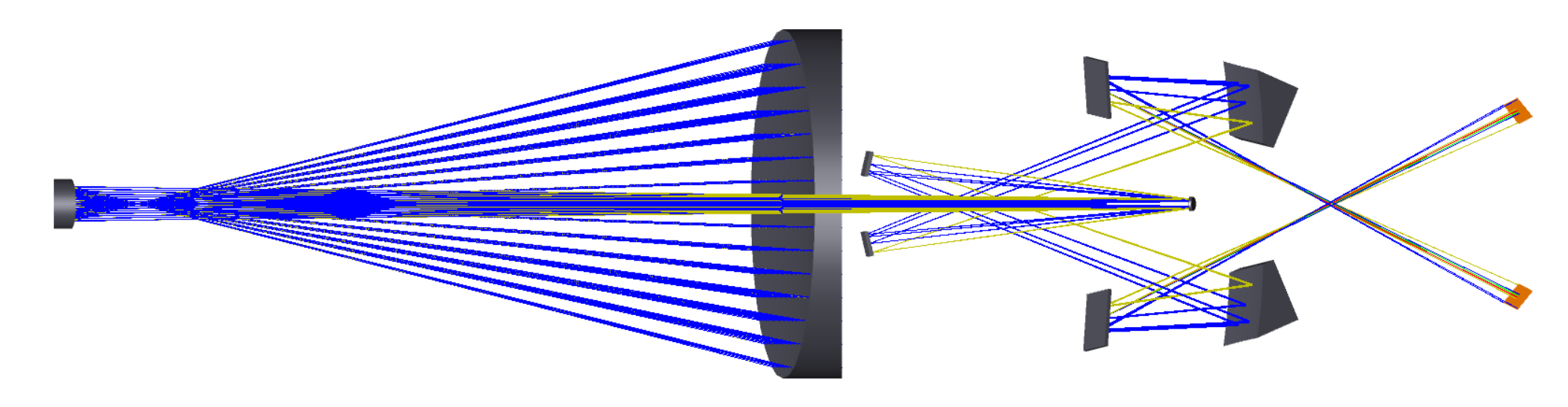}
        }
\vspace{-0.05\textwidth}   
\centerline{\Large \bf     
\hspace{0.1 \textwidth}  \color{black}{(a)}
   \hfill}
\centerline{\hspace*{0.015\textwidth}
         \includegraphics[width=1\textwidth,clip=]{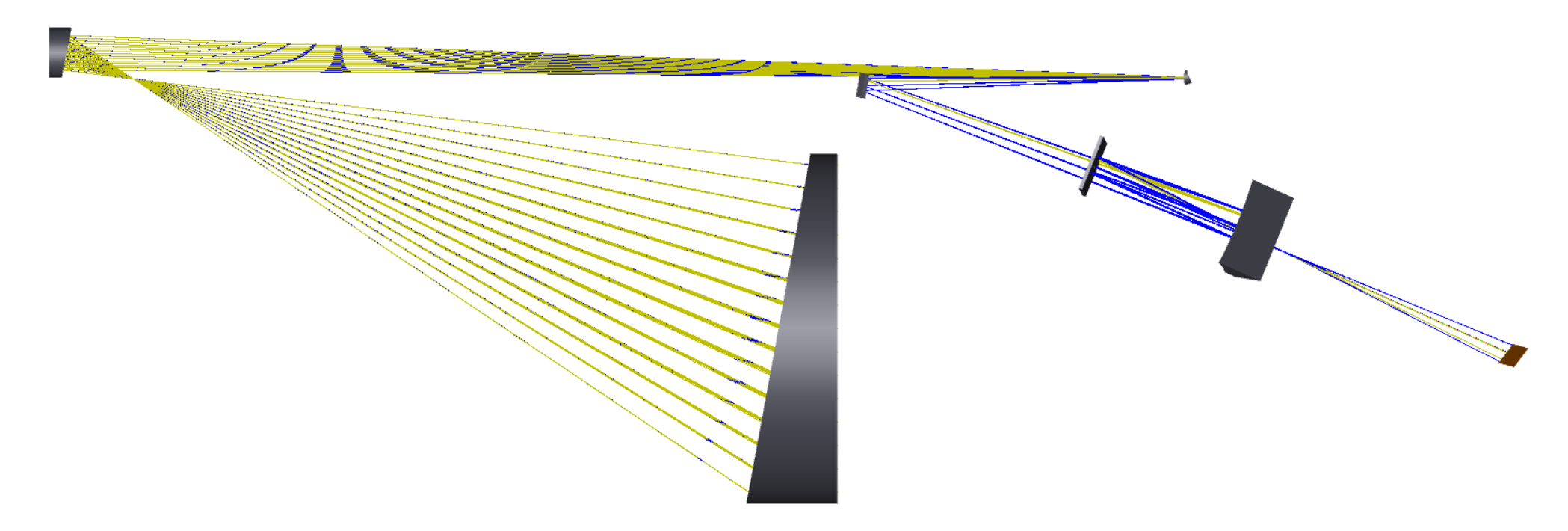}
        }
\vspace{-0.05\textwidth}   
\centerline{\Large \bf     
\hspace{0.1 \textwidth} \color{black}{(b)}
   \hfill}              
\caption{SNIFS optical design. (a) is the top view and (b) is the side view. Light enters from the left side of the image (not shown). The optical elements, in order, are: Primary Mirror, Secondary Mirror, Tertiary Mirror, Mirrorlet Arrays, Focusing Mirror, Grating, and Detector. After the third optical element, the optics split off into two different fields of view in order to view two different regions on the Sun.}
\label{F-Optics}
\end{figure}

    Each field of view of the IFS consists of a mirrorlet array, an off-axis aspheric mirror, and a flat grating. Any IFS design needs to limit its spectral bandpass to the necessary operational spectral range so that spectra from two neighboring spatial elements do not overlap significantly. SNIFS uses heritage Teledyne/Acton Optics narrowband Filters, part number FN122-N-1D.  We investigated stacking two of these filters together in order to provide sufficient out-of-band rejection, namely, of the strong visible-light components that can lead to scatter but determined that the combination of one filter and the grating is sufficient to reduce scattered light.  This filter is mounted at optical pass-through on the bulkhead from the telescope to the spectrograph sections, meaning any stray light in the telescope section would have to pass through this filter as they are the only optical path into the spectrograph section. 
    
    All non-grating mirrors are coated with Aluminum and MgF$_2$ which provides 60--80\% reflectance in the desired wavelengths, as well as high reflectance in visible light. Additional details about the optical reflectance can be found in Section \ref{S-ReflectanceTesting}.

    

\subsection{Mechanical Design}
    \label{S-MechanicalDesign}
    
The total mass of the science payload is approximately 120~kg, including the skins, bulkheads, and shutter door, and is 74.75 inches long. 

The telescope portion of the instrument is a very fast system due to the Primary Mirror's focal ratio f/\# of 1.81. Note this is different from the overall telescope's f-ratio of 273. Because of this, the Primary-Secondary relationship is extremely sensitive to misalignment, particularly thermal misalignment. Under the assumption that the skins would heat up to more than 50$^{\circ}\text{C}$, many mechanical design decisions were made to reduce the impact of thermal expansion on the secondary mirror alignment. These included making the optical benches out of stainless steel, using a 1/4$''$ thick spacer of G10 ceramic between the bulkhead and optical benches, and machining lightweighting pockets around the outer radius of the bulkhead to reduce thermal throughput along the bulkhead's radius. The launch plans were also adjusted from launching at a slight overpressure to launching at vacuum, to reduce convective heating. 

Mirror mounts were machined from titanium to have a better coefficient of thermal expansion (CTE) match with the Zerodur glass substrate of the mirrors. Fine screw adjusters were installed in the secondary and grating mounts to provide better precision control over x and y translation and z rotation. 

\subsubsection{Baffles}
    \label{S-Baffles}
\begin{figure}[!htb]
   \begin{center}
       \caption{A heat-rejection baffle was placed at the first focus of the primary mirror.This view shows the baffle mounted on a cutout of the front optics bench. The invar plate (yellow) restricts the FOV. The light grey baffle also acts as a heatsink.}
    \includegraphics[width=1\textwidth]{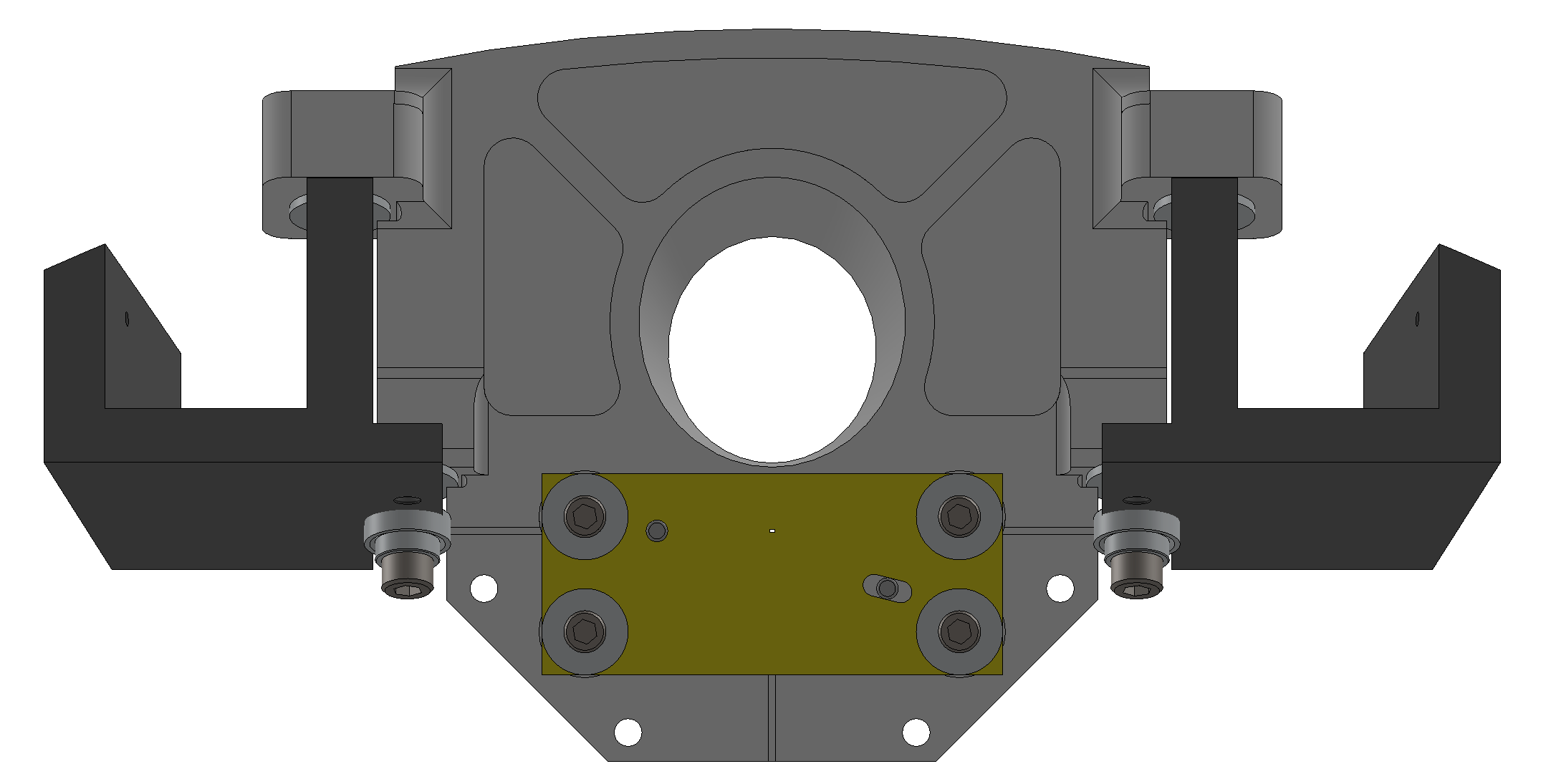}
   \end{center}
    \label{F-Tombstone}
\end{figure}

Several steps were taken to reduce heat and scattered light from reaching the spectrograph section. A 0.25$''$ thick baffle/heatsink (see Figure \ref{F-Tombstone}) was installed as a field stop at the first focus of the primary mirror normal to the incoming light. The darker plate on the front of the baffle rejects most of the focused light while a small (1~mm) hole drilled through it allows light around the field of view to pass through to the secondary mirror. This 1~mm passthrough at the first focus corresponds to a field of view approximately 8 inches in diameter at the focal plane of the mirrorlets. The larger hole in the baffle allows unfocused light from the secondary mirror to pass to the rest of the system. The remaining sunlight is projected on the surface of the baffle and imaged by the aspect camera, which will be discussed later. Over the observing duration of up to 8 minutes, the field stop surface is expected to reach a maximum temperature of 250~C due to the telescope's 95 in$^2$ of collecting area.

To reduce this thermal load transferring to the optical bench during observations, the baffle was made of stainless steel, which has poor thermal conductivity. Additionally, a 1/8 in invar plate was placed at the optical focus and thermally isolated from the rest of the baffle. The plate was bolted to the baffle and isolated using a double layer of titanium washers between the plate and the baffle, and a single layer of titanium washers between the plate and the steel screw for each of the 4 attachment points. A 1~mm FOV hole through the invar plate was cut to have a knife edge so as to reduce optical aberrations in the observations. Invar was chosen because of its low CTE, which reduced the expansion of this FOV due to heating. Finally, the baffle itself was thermally isolated from the optics bench using additional titanium washers. This design resulted in a total of five surfaces to reduce heat transfer to the optical bench. Originally, ceramic washers were planned for all thermal isolation, but they were found to have too great a risk of shattering after installation. 

In addition to the heat rejection baffle, several other tools were used to limit stray light in the spectrograph section. A sheet baffle was installed on the bottom of the front optical bench to isolate the light path after the secondary mirror from the light path before it. A Ly-$\alpha$ filter was also installed over the pass-through in the bulkhead to limit the wavelengths entering the spectrograph. See Section \ref{S-OpticalDesign} for additional information on the filter. Finally, two D-shaped cutouts about 2 inches in radius were made in the bulkhead to allow air to pass through for vacuum conductance. Two 3-inch thick air-permeable light-tight baffles were installed covering these cutouts. The baffles were made from an open-celled 6101 aluminum alloy metal foam, tradename Duocel\textsuperscript{\texttrademark}, with a porosity of 92-94 percent. After being machined into shape, the foam was conversion coated to further reduce scattered light. Due to their fragility, they were mounted in protective cases and epoxied in place before being mounted to the bulkhead. 

\subsubsection{Pointing and Attitude Determination}
    \label{S-PointingAttitude}
A coarse and fine Sun sensor provided by the Sounding Rocket Program were mounted on the front of the optical bench near the secondary mirror as per NASA requirements to provide inputs to the attitude control system. In addition, a pointing camera was installed on the front optics bench looking at the image of the Sun projected on the heat rejection baffle. This will provide in-flight knowledge and feedback of the rocket's pointing during observations. The camera is a Nocturn XL and the lens is a 50~mm fixed focal length lens from Edmund Optic. This gives us approximately a 50$\times$ magnification, which is enough to clearly distinguish the FOV hole on the invar plate as well as the edges of the Sun. Due to the brightness of the image, the lens will fly with the lens cap installed, and a 1.5~mm hole drilled through the center to avoid oversaturating the image. It was hoped that solar features such as sunspots or a flare might be visible in the image, but the current surface scatters too much light to resolve such features. Overall pointing location will be determined beforehand based on solar magnetic field maps based on research by \citealp{herde2023spicules} in order to view a location with a statistically high number of spicules.

\subsubsection{Vacuum Design}
    \label{S-VacuumDesign}

The telescope and spectrograph sections of the instrument are designed to hold a vacuum below 10$^{-4}$ torr. This is done for both cleanliness reasons and to allow for alignment and calibration using in-band EUV light. The payload will be under vacuum both during testing and while on the launch rail. Most other times, particularly during travel, the payload will be backfilled with gaseous N2 at approximately 1.1~atm to reduce the chance of external contaminants from leaking in.

The payload will launch under a vacuum between 10$^{-5}$ torr and 10$^{-3}$ torr, depending on wait times on the launch rail. After launch, the payload will be equalized with the external atmosphere once it reaches a height above 100~km and continuing until the unlatching of the shutter door. Pressure equalization is achieved by opening the solenoid valve located in the telemetry bulkhead, allowing for the exhausting of air and preventing explosive decompression when the shutter door is unlatched.

Once observations are completed, vacuum is preserved by closing both the shutter door and solenoid valve. This reduces the chance of contamination during descent and landing.

\subsection{Detector Cooling System}
    \label{S-ThermalCoolingSystem}


The SNIFS rocket utilizes two CMOS detectors (discussed in Section \ref{S-DetectorDesign}), which only require cooling to around 0 $^{\circ}\text{C}$ to reduce thermal noise.
    
Each detector is cooled by a small (0.75~kg) secondary heatsink mounted directly to it. These two secondary heatsinks are connected to a single primary heatsink with braided copper thermal straps. With a total flight duration of 10 minutes or less, active cooling of the heatsinks during flight is not required. Instead, prior to launch, the primary heatsink will be cooled to its operational temperature of -30$^{\circ}\text{C}$ to -40$^{\circ}\text{C}$, resulting in the attached detectors cooling to approximately -20$^{\circ}\text{C}$. The combination of initial temperature and mass will be sufficient to keep the detectors below 0$^{\circ}\text{C}$ for the duration of the 10 minute flight.

Only the primary heatsink is directly chilled before launch. LN2 is pumped from a dewar through an insulated flyaway line, into the heatsink, and then exhausted in gaseous form out the rocket skin. The temperature of the system is monitored continuously during the chilling process, and used to control the flow of LN2 by actuating a solenoid valve. The dewar, valve, controller, and other related equipment are mounted on the rail and remain there throughout the flight.

\subsection{Electrical Design}
    \label{S-electronics}
The SNIFS rocket electronics subsystem is comprised of FPGA boards, detectors, a power supply board, and Arduino controllers. 

SNIFS carries two detectors, one for each field of view. Each detector is paired with its own interface and FPGA boards for processing purposes as well as redundancy. These are located in the instrument's vacuum section on the underside of the optical bench.

The SNIFS electronics module provides power to the payload and command and telemetry communications to the ground. This hardware is comprised of two circuit board assemblies: the power supply board and the Arduino controller assembly. These are located in the instrument's non-vacuum section.

\subsubsection{Detector and FPGA Design}
    \label{S-DetectorDesign}

The SNIFS imager utilizes a Teledyne e2v CIS120 CMOS array sensor with a resolution of 2048$\times$2048 pixels. This sensor is back illuminated, thinned and absent of the passivation coatings that make most CMOS sensors insensitive to UV light, making it well suited for the SNIFS instrument application. It also has relatively low read noise (8e-), low dark current \citep{Otero_2021}, and a large well depth (40~000 e-/pixel). The read electronics for the detector are comprised of three boards: Detector, Interface and FPGA. Both the Interface and FPGA boards form an electronics stack that is connected to the Detector Board via a 10-inch 51-wire round-wire harness. The detector board contains a socket in which the CIS120 device is inserted. It also contains support circuitry to locally generate the numerous voltage and current references required by the device. The CIS120 is a digital output sensor with the read ADCs integrated on the die. Data are output at a rate of 260 Mbps over 4 low-voltage differential signaling pairs (LVDS) for a maximum readout rate of 20 frames per second with a pixel resolution of 12-bits. The LVDS signals utilize twisted pairs on the harness and are routed through the Interface Board to the FPGA board using impedance-controlled signals. 

The FPGA board contains a Xilinx Kintex-7 series FPGA that implements all the necessary logic to read out the detector and buffer the frame in an attached 1 GByte, DDR3 SDRAM device. A 32 GByte NAND flash chip is also placed on the board for storage of frames during the rocket flight. The NAND flash allows for the collection of a much larger amount of data than could be achieved through the rocket telemetry alone. Only one out of every seven images is expected to be downlinked during the flight, so this memory is vital for the full achievement of the mission's science goals. Due to this, recovery of the rocket is vital to achieving comprehensive mission success. The rocket telemetry interface is made up of an 8-bit, 2.5 Mbyte/s, parallel bus that is converted to LVDS pairs on the Interface board. The FPGA firmware contains a Xilinx Microblaze soft core that runs the software application which services commands and controls the data capture. Commands and housekeeping telemetry are communicated with the Arduino controller assembly using UART over RS422, with the RS422 transceiver located on the Interface Board. The Interface Board also contains 16 ADC channels for collection of voltage, current, and temperature data as well as a Real Time Clock for use by software to time-tag data.

\subsubsection{Power Supply Board}
    \label{S-PowerBoard}
The power supply board receives rocket bus or ground system power (28V nominal) and provides switchable power to two detector assemblies, the pointing camera, and the solenoid vent valve. 

These voltage supplies are implemented with efficient buck switching regulators. There is an additional DC to DC converter providing a 5-volt supply to the Arduino controller assembly. The voltage regulator outputs and the system supply current are monitored and sent to the Arduino controller assembly to be incorporated into the housekeeping telemetry.

\subsubsection{Arduino Controller Assembly}

\begin{figure}[!htb]
        \caption{SNIFS Arduino Architecture}
        \begin{center}
            \caption{Block diagram showing the commands and connections which pass through the SNIFS Arduino boards.}
    	\includegraphics[width=1\textwidth]{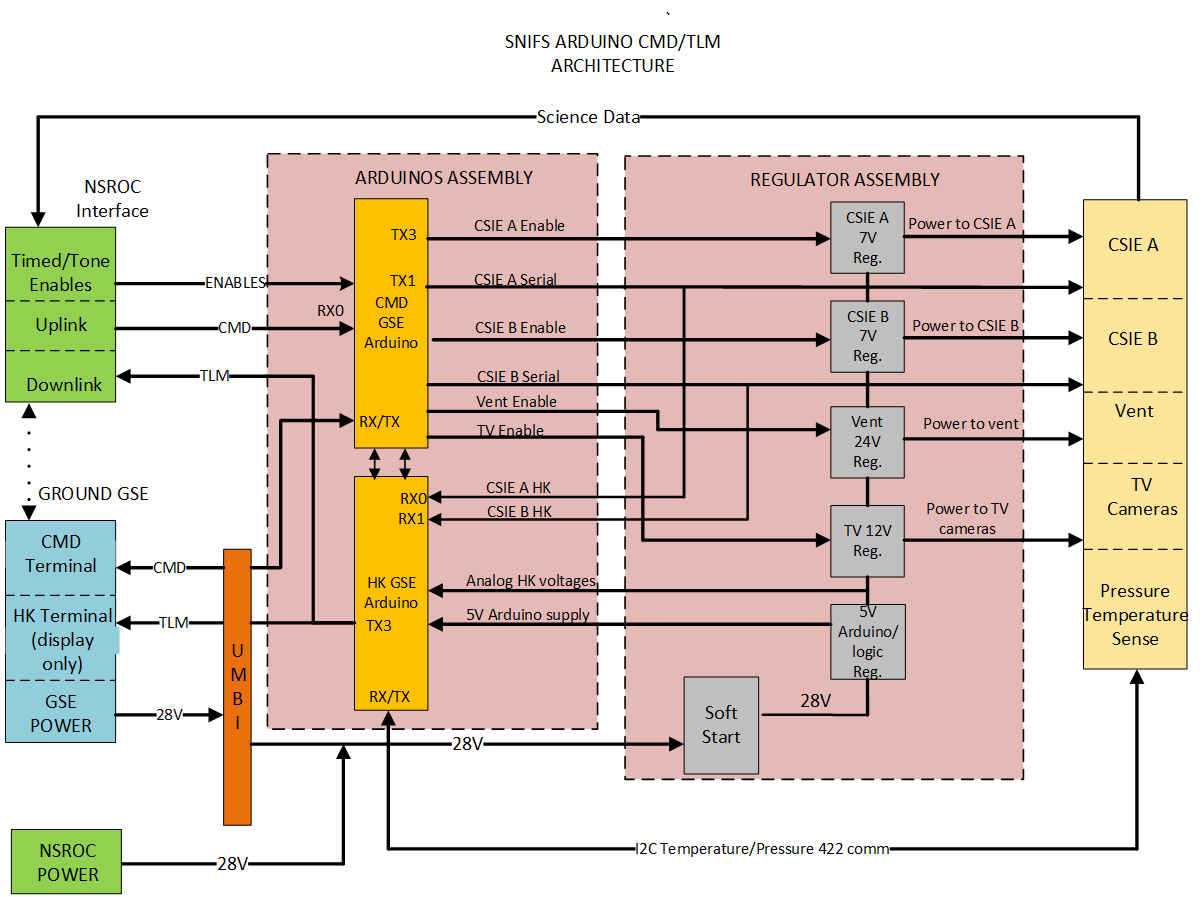}
        \end{center}
    \label{F-Arduino-plot}
    \end{figure}

This assembly contains two commercial off-the-shelf, Arduino Due microcontroller boards, based on the ATMEL ATSAM3X8E processor. Figure~\ref{F-Arduino-plot} shows a block diagram detailing these boards' connections to various subsystems.

One Arduino is dedicated to receiving data from payload temperature, pressure sensors, and status from the two detectors, and transmitting these data to the rocket telemetry interface. 

A second Arduino handles ground commands. These commands can enable or disable power to the payload subsystems. Detector command packets from the ground are received by this Arduino and relayed to the detector subsystems. All communications between the ground down and the detector modules is implemented via differential RS-422 protocols.

\subsection{Flight Software}
    \label{S-SoftwareDataProcessing}

The SNIFS flight software (FSW) is a derivative of the home-grown CU/LASP flight software package called the LASP Common Code (LCC). The LCC has been used for multiple in-house projects, especially for recent cubesats and instruments such as the \textit{Relativistic Electron and Proton Telescope integrated little experiment—2} onboard the \textit{Colorado Inner Radiation Belt Experiment} \citep[REPTile-2; CIRBE]{li2024first}, the \textit{Compact Total Irradiance Monitor} \citep[CTIM]{harber2019compact}, the \textit{Atmospheric Effects of Precipitation through Energetic X-rays} cubesat \citep[AEPEX]{marshall2023aepex}, the \textit{Climatology of Anthropogenic and Natural VLF wave Activity in Space} cubesat \citep[CANVAS]{marshall2022canvas}, the  \textit{Occultation Wave Limb Sounder} onboard the \textit{Compact SOLSTICE Experiment}  \citep[OWLS, CSOL]{thiemann2023solar}, and the \textit{Virtual Super-resolution Optics with Reconfigurable Swarms} cubesat \citep[VISORS]{sandnas2024concept}. The LCC is constantly being improved to incorporate more features, especially for the CSIE detectors.  Additionally, for SNIFS, the in-house command and control ground software, Hydra, was adapted to control the two SNIFS CSIE detectors that are now using the larger format Teledyne/e2v CIS120 chips from the VISORS CIS115 chip version. SNIFS is also running in 12-bit resolution mode from the previous 16-bit. 

The SNIFS FSW takes the image packets generated when taking data, packages them into the 8-bit parallel data required by the NASA/NSROC telemetry system and adds header information, and then transfers the data packets as requested to be downlinked to the NASA ground station and into the SNIFS computers for storage and data processing.

The SNIFS FSW is based on a Deterministic Executive Loop (DEL). The DEL implements a phase-frequency algorithm for scheduling, allowing modular software tasks to be executed independently at fixed loop iterations with collisions reported by the software. High priority tasks can be set to run on every loop iteration without collision reporting. System tasks include command and telemetry handling, command sequence execution, storage access for post-flight data recovery, and CIS configuration and control. Commands and telemetry are formatted as CCSDS packets \citep{book2003space}.

\begin{table}[!htb]
    \centering
    \begin{tabular}{p{0.25\linewidth} | p{0.6\linewidth}}
     \hline                   
         \multicolumn{2}{c}{Capture State}\\
          \hline                   
         IDLE & Does nothing\\
         INTG\textunderscore WAIT &  Waits until the integration timer has expired\\
         CAPTURE& Updates the write address used by the FPGA to place raw images into the DDR3 SDRAM\\
          \hline                   
         \multicolumn{2}{c}{Downlink State}\\
          \hline                   
         IDLE & Does nothing\\
         INIT & Writes an optional transform table to the FPGA\\
         START & Generates a metadata packet for ground processing then configures the FPGA to read the desired location from the DDR3 SDRAM and streams the resulting processed image to both NAND flash storage and the rocket telemetry interface\\
         SEND & Waits for the FPGA to finish streaming the processed image, then updates the read address used by the FPGA to retrieve raw images from the DDR3 SDRAM\\
          \hline                   
    \end{tabular}
    \caption{Flight Software Commands}
    \label{t-FlightSoftware}
\end{table}

The CIS configuration and control task (CSIE FSW) provides pre-defined preset configurations and the capability to edit any configuration parameter by command. Capture and downlink are controlled by a pair of state machines. The capture state machine includes three states while the downlink state machine includes four. These states as well as a brief description of their functions are shown in Table~\ref{t-FlightSoftware}. 

These state machines can be run independently or simultaneously. Both can be started by command or by commanding the CSIE FSW to start on the next boot, and then power cycling the system.



\section{Testing} 
    \label{S-Testing}

\subsection{Detector Quantum Efficiency}
    \label{S-DetectorQEtesting}

Three detector systems were tested in a LASP facility built to characterize and test detectors in the UV, described in detail by \citealp{Nell_2016, Nell_2021}. Due to the SNIFS operational bandpass of 120.5--122~nm, there has been no AR coating applied to any of the sensors for SNIFS, but all sensors have been passivated to enhance quantum efficiency (QE) across the ultraviolet bandpass \citep{ Jerram_2010,Heymes2020b}. At time of writing, Teledyne-e2v does not have the capability to characterize detectors at these wavelengths ($<250$ nm) and SNIFS sensors had only been tested down to 300~nm prior to delivery to LASP. There is little data published data on the QE of passivated CMOS sensors in the far ultraviolet (FUV), though some references suggest QEs as high as 23\% are possible in the FUV \citep{davisFarUltravioletSensitivity2012}. However, due to the process known to be used for these sensors the best predictions are expected to come from testing performed on CCDs with the same QE enhancement processes applied \citep{Heymes2020a, Heymes2020b}. In the
SNIFS bandpass the expectation from \citealp{Heymes2020a, Heymes2020b} is
$\sim$10\%. We note that this value accounts for quantum yield (QY),
which is not negligible in the FUV. Predictions in the SNIFS bandpass
are additionally complicated due to the absorptance of SiO$_2$ in this
bandpass. The reflectivity for one sensor was measured from 92--180~nm
where the extinction coefficient of SiO$_2$ undergoes a large
change. A model with the thickness of the SiO$_2$ layer as a free
parameter was fit to this data resulting in an estimated SiO$_2$ layer
thickness of 3.4~nm.

The available time to test sensors for SNIFS was heavily abbreviated
and there was not sufficient time to optimize all bias voltages on the
sensor to minimize issues such as charge spillback. Abbreviated
testing showed the gain to be $\sim$12 e$^-$/ADU, a value consistent
with the test sheet reports from Teledyne e2v. QE testing at near
ultraviolet (NUV) wavelengths ($\sim$250--300~nm) showed results
consistent with \citealp{Heymes2020a, Heymes2020b}. QE at NUV and visible
wavelengths is then used to estimate charge collection efficiency as a
function of the absorption length of silicon using the semi-empirical
model from \citealp{stern1994quantum}. Measurements were made at 121.6~nm and
after correcting for quantum yield resulted in a QE of 26\% in the
SNIFS bandpass. Additional testing is being performed with a spare
sensor including bias optimization and the results of those tests will
be presented in a future publication when testing is completed.

Flat field illumination tests were performed across various FUV, NUV,
and visible wavelengths. Patterns from annealing, part of the QE
enhancement passivation process, are visible at certain
wavelengths and in long dark exposures at higher temperatures. These
features are anticipated based on the methods required to passivate
the sensors and have been observed on other passivated devices
\citep{Wulser_2012, DePontieu_2014}. Flat field tests did demonstrate
notable wavelength dependence, especially at the shorter UV
wavelengths where the effects of the SiO$_2$ layer become more
prominent. Ultimately it is possible to use typical flat fielding
techniques to remove these features from images, but we note the
importance of testing in band, especially at wavelengths $<150$~nm, in
order to properly perform flat field corrections.

\subsection{Reflectance Testing}
    \label{S-ReflectanceTesting}

Four different mirrorlet arrays were manufactured and tested in a LASP facility to determine their reflectance. Each array's reflectance was tested both at its center and at one corner. Reflectances were found to range between 62--65\% at wavelengths between 121.3~nm and 121.8~nm. Reflectances were lowest at the corners of the arrays, but this may be due to the fact that the light beam overfilled the corners, resulting in efficiency loss. Due to the small radius of curvature and thus high power of all other optics, the geometry of the testing facility precluded measurements of other optics at LASP.

Reflectance testing for all other elements was provided by their manufacturers. Each mirror was found to have at least 82\% reflectance over the bandpass from 120.2--122~nm, while the grating was found to have an efficiency of 58--65\% at second-order 121.25~nm.

    
\subsection{Nominal vs Actual Resolutions}
    \label{S-Resolution}

    While the system was designed to have a theoretical spatial resolution of 0.5$''$ and a spectral resolution of 0.33~m\AA   under ideal conditions, the actual resolutions do not achieve these limits due to mechanical capabilities and narrow alignment tolerances.  The fast nature of the primary mirror requires alignment capabilities of less than one thousandth of an inch in all directions which proved difficult to achieve. Final resolution values and any additional testing will be published after the SNIFS rocket flight in Summer 2025.

\section{Conclusion} 
      \label{S-Conclusion} 
The Solar eruptioN Integral Field Spectrograph (SNIFS) instrument is a novel type of integral field spectrograph which will expand current IFS technology to new UV wavelengths. It is scheduled to launch on a sounding rocket summer 2025 from White Sands Missile Range, NM where it will take around five minutes of observations viewing the solar chromosphere and transition region. 

SNIFS is designed to have two 32$''\times32''$ fields of view each with a spatial resolution of 0.5$''$ and a spectral resolution of 33~m\AA, comparable to other solar instruments. Its spectral range from 120.5--122~nm covers three prominent chromospheric lines: Si~{\sc iii} (1206~\AA), Ly-$\alpha$ (1216~\AA), and O~{\sc v} (1218~\AA) providing observations into multiple layers of the chromosphere. SNIFS will also have a 1s observation cadence which is an order of magnitude faster than current similar capabilities.

SNIFS will be able to observe many solar chromospheric and transition region phenomena, but will excel at observing the structure of fast-changing events due to its high time cadence. Such events include solar flares, nanoflares, and solar spicules. Its observations will be used as constraints for RADYN simulations to further our understanding of mass and energy transport between the solar surface and corona and provide another building block in in the field's work understanding the coronal heating problem. 

In addition to its science mission, SNIFS will help to develop space technologies in several ways. It is the first in-space demonstration of a new integral field spectrograph design which utilizes mirrorlet arrays to take UV spectral images of a region of the Sun which is notoriously hard to study. It also advances the development of a new CMOS detector design. Finally, SNIFS has proved a training ground for new scientists and engineers, providing training for five undergraduate students and two graduate students during its development, assembly, and flight.

\begin{acknowledgments}
  V. L. H., P. C. C., D. S., A. D., V. P., and S.B. would like to acknowledge support from NASA Grant 80NSSC22K0576 (SNIFS, PI: Chamberlin) and the NASA H-FORT program. 
  
  S.B. and V.P. also gratefully acknowledge support from NASA contract \\ NNG09FA40C (IRIS) to LMSAL.
  
  R.O.M. would like to thank the Science and Technologies Facilities Council (UK) for the award of an Ernest Rutherford Fellowship (ST/N004981/2) and a New Applicant grant (ST/W001144/1), as well as NASA for support from a heliophysics supporting research grant (NNH19ZDA001N-HSR/80NSSC21K0010).

  T.V would like to thank Beth Cervelli (LASP-FSW) for design and implementation of the LASP Common Code Flight Software package and the Hydra Command and Control Ground Software system.
\end{acknowledgments}


\bibliographystyle{spr-mp-sola}
\bibliography{SNIFS_Main_body}

\end{document}